\documentclass[12pt,preprint]{aastex}

\shorttitle{GRMHD Simulations of Jet Formation} \shortauthors{Mizuno
et al.}

\usepackage{graphicx}

\begin{document}

\title{General Relativistic Magnetohydrodynamic Simulations of \\
Jet Formation with a Thin Keplerian Disk}

\author{
Yosuke Mizuno\altaffilmark{1,6}, Ken-Ichi
Nishikawa\altaffilmark{1,2}, Shinji Koide\altaffilmark{3}, Philip
Hardee\altaffilmark{4} \\ and Gerald J. Fishman\altaffilmark{5} }

\altaffiltext{1}{National Space Science and Technology Center, 320
Sparkman Drive, VP 62, Huntsville, AL 35805, USA;
Yosuke.Mizuno@msfc.nasa.gov} 
\altaffiltext{2}{Center for Space
Plasma and Aeronomic Research, University of Alabama in Huntsville}
\altaffiltext{3}{Department of Physics, Kumamoto University,
Kurokami, Kumamoto, 860-8555, Japan} 
\altaffiltext{4}{Department of
Physics and Astronomy, The University of Alabama, Tuscaloosa, AL
35487, USA}
\altaffiltext{5}{NASA-Marshall Space Flight Center,
National Space Science and Technology Center, 320 Sparkman Drive, VP
62, Huntsville, AL 35805, USA} 
\altaffiltext{6}{NASA Postdoctoral
Program Fellow/ NASA Marshall Space Flight Center}

\begin{abstract}

We have performed several simulations of black hole systems
(non-rotating, black hole spin parameter $a=0.0$ and rapidly
rotating, $a=0.95$) with a geometrically thin Keplerian disk using
the newly developed RAISHIN code. The simulation results show the
formation of jets driven by the Lorentz force and the gas pressure
gradient. The jets have mildly relativistic speed ($\gtrsim 0.4~c$).
The matter is continuously supplied from the accretion disk and the
jet propagates outward until each applicable terminal simulation
time (non-rotating: $t/\tau_{\rm S} = 275$ and rotating:
$t/\tau_{\rm S} = 200$, $\tau_{\rm S} \equiv r_{\rm S}/c$). It
appears that a rotating black hole creates an additional, faster,
and more collimated matter-dominated inner outflow ($\gtrsim
0.5~c$) formed and accelerated by the twisted magnetic field
resulting from frame-dragging in the black hole ergosphere.
This is the first known simulation confirming the formation of an inner
magnetically-driven, matter-dominated jet by the frame-dragging
effect from a black hole co-rotating with a thin
Keplerian disk threaded by a vertical magnetic field. This
result indicates that jet kinematic structure depends on black hole
rotation and on the initial magnetic field configuration and
strength.

\end{abstract}

\section{Introduction}

Both magnetic and gravitational fields play an important role in the
dynamics of matter in many astrophysical systems. In a highly
conducting plasma, the magnetic field can be amplified by gas
contraction or shear motion. Even when the magnetic field is weak
initially, the magnetic field can grow on short time scales and
influence the gas dynamics of the system. The magnetic influence on
gas dynamics is particulary important for a compact object such as a
black hole or a neutron star. Relativistic jets have been observed
or postulated in various astrophysical objects, including active
galactic nuclei (AGNs) (e.g., Urry \& Pavovani 1995; Ferrari 1998),
microquasars in our galaxy (e.g., Mirabel \& Rodiriguez 1999), and
gamma-ray bursts (GRBs) (e.g., Zhang \& M\'{e}sz\'{a}ros 2004; Piran
2005; M\'{e}sz\'{a}ros 2006). The most promising mechanisms for
producing the relativistic jets involve magnetohydrodynamic
centrifugal acceleration and/or magnetic pressure driven
acceleration from an accretion disk around the compact objects
(e.g., Blandford \& Payne 1982; Fukue 1990), or involve the extraction of
rotating energy from a rotating black hole (Penrose 1969; Blandford
\& Znajek 1977).

General relativistic magnetohydrodynamics (GRMHD) codes with
fixed spacetimes have been developed to investigate relativistic magnetorotators
(RMRs) (e.g., Koide et al. 1998; De Villiers \& Hawley 2003; Gammie et al.\ 2003;
Komissarov 2004; Ant\'{o}n et al.\  2005; Anninos et al.\ 2006).
These codes have been used to study the Blandford-Znajek
effect near a rotating black hole (Koide 2003; Komissarov 2005;
McKinney 2005), and the formation of GRB jets in collapsars (Mizuno
et al.\ 2004a, 2004b; De Villiers et al.\ 2005b).

Recently, in order to investigate the properties of accretion flows
onto a black hole associated with the magneto-rotational instability
(MRI) (Balbus \& Hawley 1991), many simulations have been performed
using a thick torus-like disk (the disk thickness $H/r > 0.1$, where
$H$ is height of the disk and $r$ is radius from a black hole) with
weak poloidal magnetic fields in a torus (the plasma beta, $\beta =
p_{\rm gas}/p_{\rm mag} > 100$, where $p_{\rm gas}$ is gas pressure
and $p_{\rm mag}$ is magnetic pressure) (De Villiers et al.\ 2003,
2005a; Krolik et al.\ 2005; Hawley \& Krolik 2006; Beckwith et al.\
2006; McKinney \& Gammie 2004; McKinney 2006). The initial
``poloidal-loop'' magnetic fields in the torus contribute to the
generation of MRI, diffusion of matter and magnetic field, and jet
generation. However, in their simulations the structure of magnetic
fields that piled up and twisted near the black hole is different from
the magnetic fields which are twisted by a thin Keplerian disk
and/or the frame-dragging effect of the rotating black hole ($H/r
\sim 0.06$) with a stronger initial vertical magnetic field ($\beta
< 40$) (Koide, Shibata, \& Kudoh 1998, 1999; Koide et al. 2000;
Nishikawa et al. 2005). In the thin disk simulations MRI does not
grow since the wavelength at the maximum growth rate is larger
than the height of the thin disk. Koide et al.\ (2000) have found
that jets are formed from thin Keplerian accretion disks for both
counter- and co-rotating black holes. In the co-rotating disk case, 
the jet had a two-layered structure: an inner gas pressure-driven jet, 
and an outer magnetically driven jet (Koide et al. 2000).

In this Letter we report on new simulations showing jet formation
from the black hole magnetosphere co-rotating with a thin Keplerian
disk threaded by a vertical magnetic field using the recently
developed three-dimensional, GRMHD code RAISHIN (Mizuno et al.
2006).  Our new result shows
that an inner matter-dominated jet is generated by the
magnetic field twisted in the ergosphere of the rotating black hole
in addition to the outer jet formed by the co-rotating accretion
disk seen by Koide et al.\ (2000). While our inner jet is similar
the inner jet created by a counter-rotating disk found by Koide et al.
(2000), in this case the inner jet is located closer to the rotation
axis.

\section{Numerical Method}

In order to study the formation of relativistic jets from a geometrically thin
Keplerian disk, we use a 2.5-dimensional GRMHD code with Boyer-Lindquist
coordinates $(r, \theta, \phi)$. The method is based on a 3+1 formalism of the
general relativistic conservation laws of particle number and energy momentum,
Maxwell equations, and Ohm's law with no electrical resistance 
(ideal MHD condition) in a curved spacetime (Mizuno et al. 2006).

In the RAISHIN code, a conservative, high-resolution shock-capturing scheme is
employed. The numerical fluxes are calculated using the Harten, Lax, \& van Lee
(HLL) approximate Riemann solver scheme. The flux-interpolated, constrained
transport scheme is used to maintain a divergence-free magnetic field. The
RAISHIN code has proven to be accurate to second order and has passed numerical
tests including highly relativistic cases, and highly magnetized cases in both
special and general relativity.  Code details and the results of code tests can
be found in Mizuno et al.\ (2006). In the simulations presented here we use
minmod slope limiter reconstruction, HLL approximate Riemann solver, flux-CT
scheme and Noble's 2D method.

In our present simulations: a geometrically thin Keplerian disk
rotates around a black hole (non-rotating, $a=0.0$ or rapidly
rotating, $a=0.95$, here $a$ is black hole spin parameter), where
the disk density is 100 times higher than the coronal density. The
thickness of the disk is $H/r \sim 0.06$ at $r=10 r_{\rm S}$. In the
rotating black hole case the disk is co-rotating with the black
hole. The background corona is free-falling into the black hole
(Bondi flow). The initial magnetic field is assumed to be uniform
and parallel to the rotational axis i.e., the Wald solution (Wald
1974). Our scale-free simulations are normalized by the speed of
light, $c$, and the Schwarzschild radius, $r_{\rm S}$, with
timescale, $\tau_{\rm S} \equiv r_{\rm S}/c$. Values of the magnetic
field strength and gas pressure depend on the normalized density,
$\rho_{0}$. In these simulations the magnetic field strength,
$B_{0}$, is set to $0.05 \sqrt{\rho_{0} c^{2}})$. These initial
conditions are similar to Koide et al.\ (1999, 2000) but with a
weaker initial magnetic field. Koide et al.\ (2000) used $B_{0}=0.3
\sqrt{\rho_{0} c^{2}}$. The simulations are performed in the region
$1.1 r_{\rm S} \le r \le 20.0 r_{\rm S}$ (non-rotating black hole
case) and $0.75 r_{\rm S}\le r \le 20.0 r_{\rm S}$ (rapidly rotating
black hole case) and  $0.03 \le \theta \le \pi/2$.  We use $128
\times 128$ computational zones with logarithmic zone spacing in the
radial direction (Koide et al.\ 1999). We assume axisymmetry with
respect to the $z$-axis and mirror symmetry with respect to the
equatorial plane. We employ a free boundary condition at the inner
and outer boundaries in the radial direction through which waves,
fluids, and magnetic fields can pass freely.

In our simulations, $\beta$ is higher than 1 in the disk initially
($\beta \sim 40$ at $r=5r_{\rm S}$). Therefore, in principle, MRI
grows. However, the wavelength at the maximum growth rate is
$\lambda_{\rm MRI} \sim v_{\rm Az}/\Omega $ where $\Omega$ is the
angular velocity and $v_{\rm Az}$ is the $z$-component of the
Alfv\'{e}n velocity (e.g., Balbus \& Hawley 1991, Gammie 2004). With
our initial conditions, the wavelength at the maximum growth rate is
larger than $0.6 r_{\rm S}$ at $r= 5.0 r_{\rm S}$ and larger than
the disk height ($0.3 r_{\rm S}$). Therefore the thin disk does not
support the growth of MRI at this wavelength. Also, shorter wavelengths
associated with MRI have slower growth rate and may not grow
significantly on our simulation time scale with our grid resolution.
Thus, our simulations investigate the physics of jet formation and
the properties of the jets in the context of a black hole with a
thin Keplerian disk threaded by vertical magnetic fields (using Wald
solution) in the absence of significant MRI.

In contrast, previous GRMHD simulations with a thick torus
have been used to study disk evolution via MRI and the
resulting outflows (e.g., De Villiers et al. 2003; McKinney \&
Gammie 2004; McKinney 2006; De Villiers et al. 2005; Hawley \&
Krolik 2006). Here MRI creates turbulent structure in the thick disk,
fluctuating accretion into the black hole, and unsteady
outflows near the funnel wall, e.g., Hawley \&
Krolik (2006).

\section{Results}

\begin{figure}
\epsscale{0.70} \plotone{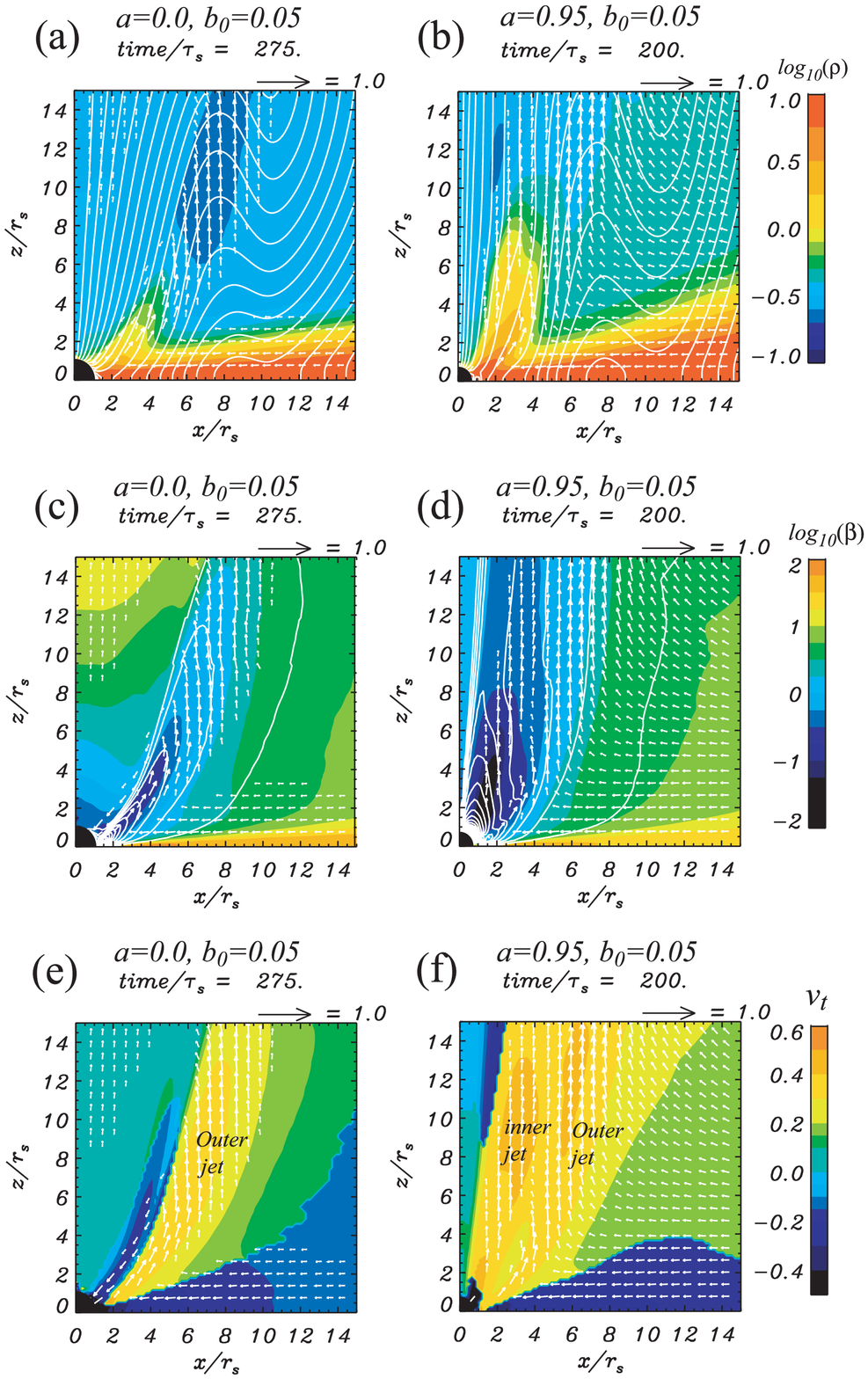} \caption{Snapshots of density and
plasma beta of the non-rotating black hole case ($a=0.0$; {\it a, c,
e}) and the rapidly rotating black hole case ($a=0.95$; {\it b, d,
f}) at the applicable terminal simulation time. The color scales
show the logarithm of density (upper panels), plasma beta ($\beta =
p_{\rm gas} / p_{\rm mag}$; middle panels) and total velocity (lower
panels). A negative velocity indicates inflow towards the black
hole. The white lines indicate magnetic field lines (contour of the
poloidal vector potential; upper panels) and contours of the toroidal
magnetic field stength (middle panels). Arrows depict the poloidal
velocities normalized to light speed, as indicated above each panel
by the arrow. \label{f1a}}
\end{figure}

Figure \ref{f1a} shows snapshots of the density (panels (a) and
(b)), plasma beta ($\beta=p_{\rm gas}/p_{\rm mag}$) distribution
(panels (c) and (d)), and total velocity (panels (e) and (f)) for
the non-rotating black hole case, $a=0.0$ (left panels); and the
rapidly rotating black hole case, $a=0.95$ (right panels); at each
simulation's terminal time (non-rotating: $t = 275\tau_{\rm S}$ and
rotating:  $t = 200\tau_{\rm S}$). The Keplerian disk at the
marginally stable circular orbit ($r = 3r_{\rm S}$) rotates around
the black hole in about $40 \tau_{\rm S}$. In the non-rotating black
hole case about 7 inner disk rotations occurred during the
simulation.

The numerical results show that matter in the disk loses angular
momentum to the magnetic field and falls into the black hole. A
centrifugal barrier decelerates the falling matter and produces a shock
around $r=2 r_{\rm S}$. Matter near the shock region is
accelerated by the $\mathbf{J} \times \mathbf{B}$ force and the gas
pressure gradient in the z-direction (see the arrows in the panels in
Figure 1). In the simulations, matter is continuously supplied from the accretion disk
and the jet propagates outward through the outer boundary of the
grid. In general, the results are similar to the previous work
of Koide et al.\ (1998, 1999, 2000) and Nishikawa et al.\
(2005).

Contours of the toroidal magnetic field strength are shown by the
white lines in Figs.\ 1c and 1d and indicate where the magnetic field
is most twisted.
In the non-rotating black hole case the magnetic field is twisted by
the rotation of the Keplerian disk near the black hole region,
propagates outwards along the poloidal magnetic field as an Alfv\'{e}n
wave and forms a jet. In the rapidly
rotating black hole case the magnetic field is strongly twisted by
the frame-dragging effect of the rotating black hole near the black hole
region e.g., Figure 1d contours and Figure 2, propagates outwards as an 
Alfv\'{e}n wave along the poloidal magentic field lines shown in Fig.\ 1b
 and forms an additional inner jet component closer to the black hole along the
rotation axis.

The total velocity distribution of non-rotating and rapidly rotating
black hole cases are shown in Figs. 1e and 1f. The jets in both
cases have speeds greater than $0.4 c$ (mildly relativistic) and the
speeds are comparable to the Alfv\'{e}n speeds. In the jets,
toroidal velocity is the dominant velocity component (see Figs. 3a
and 3d). In the rapidly rotating black hole case the velocity
distribution indicates a two-component jet. The outer jet is similar to
that of the non-rotating black hole case but the inner jet is not
seen in the non-rotating black hole case. The inner jet is faster
than the outer jet (over $0.5c$).

 Figure 3 shows the distribution of various
physical quantities on the $z/r_{\rm S} = 2$ surface at the terminal
simulation time and allows us to examine the relationship between
the twisted magnetic field and the jet velocity.  In the
non-rotating black hole case the jet is generated in the region $r
\ge 2.7 r_{\rm S}$ as shown in Figs.\ 1a and 3a. In this region, the
toroidal velocity ($v_{\phi}$) is the dominant velocity component.
The  magnetic field is twisted by the rotation of the Keplerian
disk. As the jet propagates outward, the magnetic field is
concentrated towards the rotation axis and the vertical component of
the magnetic field, $B_{\rm z}$, becomes larger near the rotation
axis. In the rapidly rotating black hole case, a jet is generated in
the region $r \ge 0.5 r_{\rm S}$. Now, two jet components
 can be seen in the velocity distribution. An inner jet is located
at $0.7 r_{\rm S} \le r \le 2.0 r_{\rm S} $ and an outer jet is
located at $r \ge 2.0 r_{\rm S}$ (Fig.\ 3d). For both jet
components, the toroidal velocity component is dominant. The
toroidal velocity of the inner jet is larger than that of the outer
jet. Here the magnetic field near the black hole region is
strongly twisted by the frame-dragging effect as shown in Fig. 2.
Since the magnetic field is twisted and pinched toward the rotation
axis, the $z$ component of the magnetic field becomes large near the
rotation axis (Fig.\ 3e), and the inner jet accelerates more rapidly
than the outer jet.

To confirm the jet acceleration mechanism, we evaluate the vertical
components of the Lorentz force, $\mathbf{F}_{\rm EM}= \rho_{\rm e}
\mathbf{E} + \mathbf{J} \times \mathbf{B}$ and the gas pressure
gradient, $\mathbf{F}_{\rm gp} = - \nabla p$ at the $z=2.0 r_{\rm
S}$ surface (e.g., Figs.\ 3c and 3f). Although this analysis is
simpler than that of Hawley \& Krolik (2006), it illustrates which 
force dominates jet formation, and has been used in previous work 
(e.g., Koide et al 1999, 2000, 2006; Mizuno et al.
2004a,b; Nishikawa et al. 2005).

This simple analysis clearly shows that the outer jet is accelerated mainly by
the gas pressure gradient. The inner part of the outer jet may be
accelerated partially by the Lorentz force but the Lorentz force in the
outer part of the outer jet in the rapidly rotating black hole case is
lower than the gas pressure gradient by an order of magnitude. The inner jet
is accelerated by the Lorentz force. Therefore the acceleration
mechanism is different in the inner and outer jets.

\section{Summary and Discussion}

We have performed simulations of jet formation from a geometrically
thin accretion disk near both non-rotating and rotating black holes
using the newly developed RAISHIN code (Mizuno et al. 2006).
Simulation results show the formation of magnetically-driven jets
accelerated by the Lorentz force and the gas pressure. In general, results
are similar to those found in previous GRMHD
simulations with a thin Keplerian disk (Koide et al.\ 1998, 1999,
2000; Nishikawa et al. 2005). However, in this study the rotating
black hole creates a second, faster, and more collimated inner
matter-dominated jet formed by the twisted magnetic field resulting
from frame-dragging in the black hole ergosphere. Thus, kinematic
jet structure depends on black hole rotation. While an inner jet is
created by a counter-rotating thin
Keplerian disk (Koide et al. 2000), the inner jet seen in
our simulations is a new result not found in a previous co-rotating 
thin Keplerian disk case (Koide et al.\ 2000). This difference appears to
be the result of the weaker initial magnetic field strength used in
the present simulations. With stronger magnetic field (Koide et al.\ 2000),
transport of angular momentum from the disk is more rapid, 
accreting matter falls
more quickly, is decelerated by the centrifugal
barrier near the black hole strongly, and generates a stronger shock.
As a result, jets are accelerated by the gas-pressure rather than
the Lorentz force of magnetic field twisted by frame-dragging. The
jet seen in the co-rotating disk simulation in Koide et al. (2000)
corresponds to the outer jet seen in our present co-rotating disk
simulation. Therefore this is a first simulation confirming the
formation of an inner matter-dominated jet driven by the magnetic
field twisted by the frame-dragging effect from a black hole
co-rotating with a thin Keplerian disk threaded by a vertical
magnetic field.

A two component jet structure has also been seen in GRMHD
simulations of a black hole co-rotating with a thick torus (e.g.,
Hawley \& Krolik 2006; McKinney 2006). One component is a
matter-dominated outflow (funnel wall jet) with a mildly
relativistic velocity ($\sim 0.3c$) along the centrifugal barrier
accelerated and collimated by magnetic and gas pressure forces in
the inner torus and the surrounding corona. This formation mechanism
and jet are the same as the outer jet seen in our simulations.
Therefore the funnel wall jet in Hawley \& Krolik (2006) is the same
as the outer jet in our GRMHD simulations. Their other component is
a highly-relativistic Poynting flux dominated jet that is produced
from the formation of a large scale radial magnetic field within the
funnel. In our simulation, such a highly-relativistic Poynting flux
dominated jet is not seen. This is likely caused by the difference
in the initial magnetic field configuration. In Hawley \& Krolik
(2006) and McKinney (2006), there are initial ``poloidal-loop''
magnetic fields inside the torus. In the simulations, the magnetic field is
twisted and expands from the torus as a magnetic tower and fills the
funnel region (Hirose et al. 2004). In the late stage of the
simulations a highly-relativistic Poynting flux dominated jet is
formed in the low density regions in the funnel. On the other hand,
our magnetic field vertically threads the disk and ergosphere
initially. The vertical magnetic field near the black hole region is
twisted by the frame-dragging effect (see Fig. 2) and forms the
inner jet. The inner jet is much denser than the highly-relativistic
Poynting flux dominated jet of Hawley \& Krolik (2006) and McKinney
(2006) because in the inner jet the matter is supplied from the
accretion disk and the free-falling corona. While our jets are
formed by the same basic mechanisms as in Hawley \& Krolik (2006)
and McKinney (2006), the different initial magnetic configuration
has led to different jet properties and a slower matter dominated
jet spine. Previous results with the co-rotating disk case of Koide
et al. (2000) found that a stronger magnetic field suppressed the
formation of the presently observed matter-dominated jet spine.

\acknowledgments

Y. M. is a NASA Postdoctoral Program fellow at NASA Marshall Space
Flight Center. He thanks G.\ Richardson, D.\ Hartmann, C.\ Fendt, 
and M.\ Camenzind for useful discussions. K.\ N. acknowledges 
partial support by National Science Foundation awards ATM-0100997, 
INT-9981508, and
AST-0506719, and the National Aeronautic and Space Administration
award NNG05GK73G to the University of Alabama in Huntsville. P.\
H. acknowledges partial support by National Space Science and
Technology Center (NSSTC/NASA) cooperative agreement NCC8-256 and
NSF award AST-0506666 to the University of Alabama. The simulations
have been performed on the IBM p690 at the National Center for
Supercomputing Applications (NCSA) which is supported by the NSF and
Altix3700 BX2 at YITP in Kyoto University.

\newpage

\begin{figure}
\epsscale{0.7} \plotone{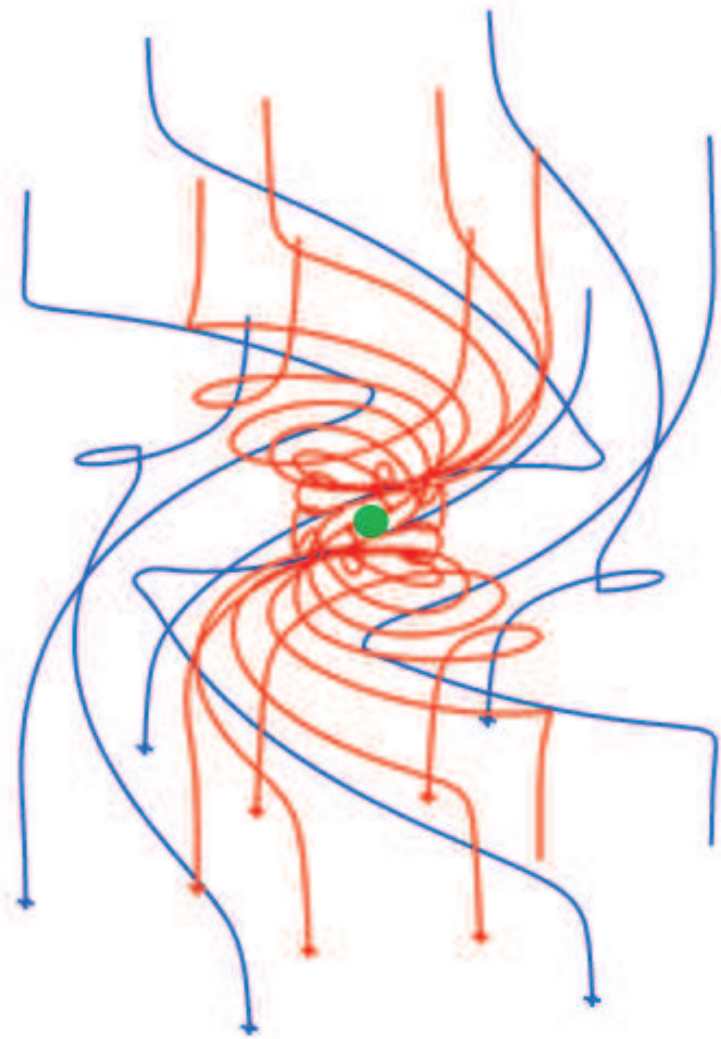} \caption{Bent and twisted magnetic
field lines obtained from the co-rotating black hole case at $t=133 
\tau_{\rm S}$. 
The field lines are traced from $z = - 15r_{\rm S}$
to $z = + 15r_{\rm S}$ beginning at  $r = 6 r_{\rm S}$ (red lines)
and  $r = 12 r_{\rm S}$ (blue lines). The black hole is indicated by 
the green circle.}
\end{figure}

\newpage

\begin{figure}
\epsscale{0.85} \plotone{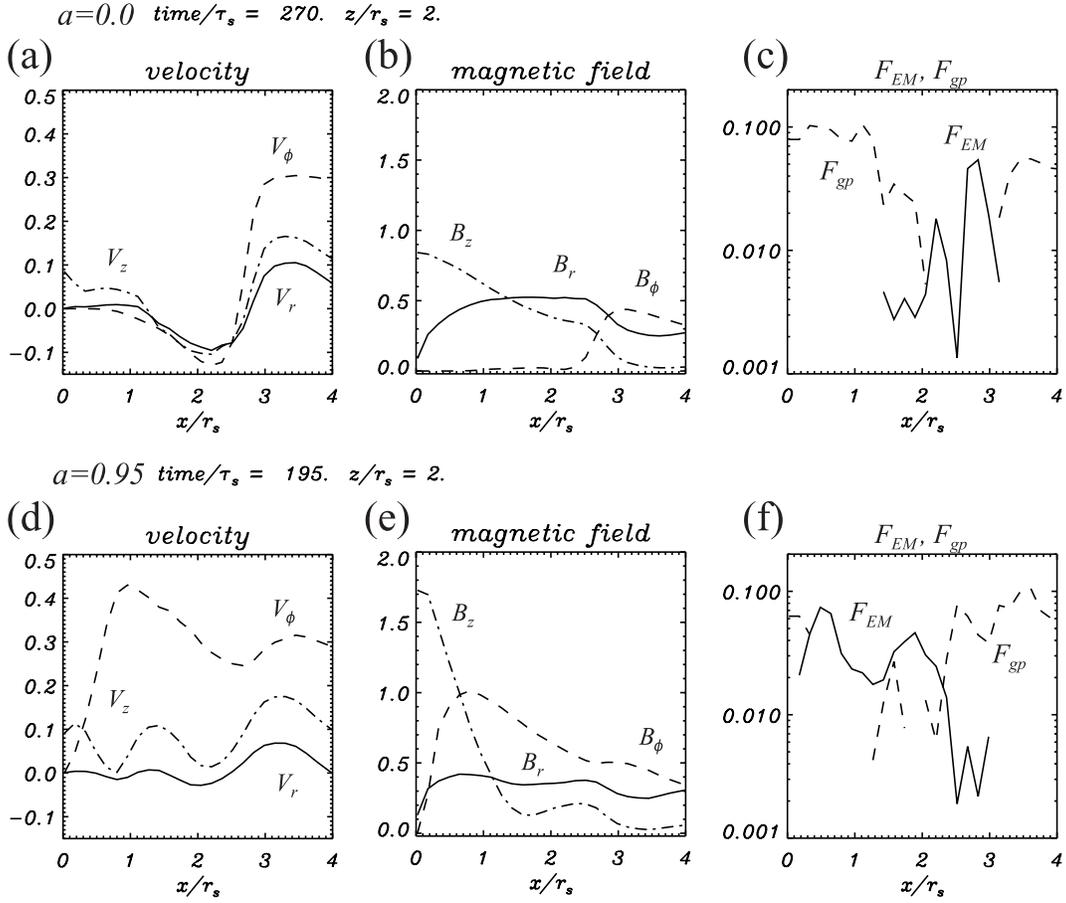} \caption{Various physical
quantities on the $z/r_{\rm S} = 2$ surface at the applicable
terminal simulation time. Panels {\it (a), (d)} show the velocity
components $v_{\rm r}$ ({\it solid line}), $v_{\phi}$ ({\it dashed
line}), and $v_{\rm z}$ ({\it dot-dashed line}). Panels {\it (b),
(e)} show the magnetic field components $B_{\rm r}$ ({\it solid
line}), $B_{\phi}$ ({\it dashed line}), and $B_{\rm z}$. ({\it
dot-dashed line}). Panels {\it (c), (f)} show the $z$ component of
the Lorentz force $\mathbf{F}_{\rm EM}$ ({\it solid line}) and the
gas pressure gradient $\mathbf{F}_{\rm gp}$ ({\it dashed line}).
\label{f3}}
\end{figure}

\newpage


\begin{thebibliography}{}

\bibitem[Anninos et al.(2005)]{Ann05} Anninos, P., Fragile, P. C., \& Salmonson,
J. D. 2005, \apj, 635, 723

\bibitem[Ant\'{o}n et al.(2006)]{Ant06} Ant\'{o}n, L., et al. 2006,
      \apj, 637, 296

\bibitem[Balbus \& Hawley(1991)]{Balb91} Balbus, S. A. \& Hawley, J. F. 1991,
\apj,376, 214

\bibitem[Beckwith et al.(2006)]{Bec06} Beckwith, K., Hawley, J. F., \& Krolik, J. H. 2006, \apj, submitted (astro-ph/0605295)

\bibitem[Blandford \& Znajek(1977)]{Bla77} Blandford, R. D. \& Znajek, R. L.
1977, \mnras, 179, 433

\bibitem[Blandford \& Payne(1982)]{Bla82} Blandford, R. D. \& Payne, D. G. 1982,
\mnras, 199, 883

\bibitem[De Villiers \& Hawley(2003)]{DeVH03} De Villiers, J.-P. \& Hawley, J. F. 2003,
\apj, 589, 458

\bibitem[De Villiers et al.(2003)]{DeV03} De Villiers, J.-P., Hawley, J. F.,
\& Krolik, J. H. 2003, \apj, 599, 1238

\bibitem[De Villiers et al.(2005a)]{DeV05a} De Villiers, J.-P., Hawley, J. F.,
Krolik, J. H., \& Hirose, S. 2005a, \apj, 620, 878

\bibitem[De Villiers et al.(2005b)]{DeV05b} De Villiers, J.-P., Staff, J., \&
Ouyed, R. 2005b, in preparation (astro-ph/0502225)

\bibitem[Ferrari(1998)]{Fer98} Ferrari, A. 1998, \araa, 36, 539

\bibitem[Gammie et al.(2003)]{Gam03} Gammie, C. F., McKinney, J. C., \& T\'{o}th, G. 2003, \apj, 589, 444

\bibitem[Gammie(2004)]{Gam04} Gammie, C. F. 2004, \apj, 614, 309

\bibitem[Fukue(1990)]{Fuk90} Fukue, J. 1990, \pasj, 42, 793

\bibitem[Hawley \& Krolik(2006)]{Haw06} Hawley \& Krolik 2006, \apj, 614, 103

\bibitem[Hirose et al.(2004)]{hiro04} Hirose, S., Krolik, J., De Villiers, J. P., \&
Hawley, J. F. 2004, ApJ, 606, 1083

\bibitem[Koide et al.(1998)]{Koi98} Koide, S., Shibata, K.,
    \& Kudoh, T.  1998, \apjl,  495,  L63

\bibitem[Koide et al.(1999)]{Koi99} -----.  1999, \apj,  522,  727

\bibitem[Koide et al.(2000)]{Koi00} Koide, S., Meier, D. L., Shibata, K.,
     \& Kudoh, T. 2000, \apj,  536,  668

\bibitem[Koide(2003)]{Koi03} Koide, S. 2003, \prd, 67, 104010

\bibitem[Koide et al.(2006)]{Koi06} Koide, S., Kudoh, T., \& Shibata, K. 2006, \prd, 74,
   044005

\bibitem[Komissarov(2004)]{Kom04} Komissarov, S. S. 2004, \mnras, 350,
      1431

\bibitem[Komissarov(2005)]{Kom05} -----. 2005, \mnras, 359, 801

\bibitem[Krolik et al.(2005)]{Kro05} Krolik, J. H., Hawley, J. F., \& Hirose, S. 2005,
   \apj, 622, 1008

\bibitem[McKinney(2005)]{McK05} McKinney, J. C. 2005, \apjl, L5

\bibitem[McKinney(2006)]{McK06} -----. 2006, \mnras, 368, 1561

\bibitem[McKinney \& Gammie(2004)]{McK04} McKinney, J. C. \& Gammie,
     C. F. 2004, \apj, 977

\bibitem[M\'{e}sz\'{a}ros(2006)]{Mez06} M\'{e}sz\'{a}ros, P. 2006, Rep. Prog.
     Phys., in press (astro-ph/0605208)

\bibitem[Mirabel \& Rodr\'{i}guez(1999)]{Mir99} Mirabel, I. F. \& Rodr\'{i}guez,
L. F. 1999, \araa, 37, 409

\bibitem[Mizuno et al.(2004a)]{Miz04a} Mizuno, Y., Yamada, S., Koide, S.,
      \& Shibata, K. 2004a, \apj, 606, 395

\bibitem[Mizuno et al.(2004b)]{Miz04b} -----. 2004b, \apj, 615, 389

\bibitem[Mizuno et al.(2006a)]{Miz06} Mizuno, Y., Nishikawa, K.-I., Koide, S.,
      Hardee, P., \& Fishman, G. J. 2006, \apjs, submitted (astro-ph/0609004)

\bibitem[Nishikawa et al.(2005)]{Nis05} Nishikawa, K.-I., Richardson, G., Koide, S.,
Shibata, K., Kudoh, T., Hardee, P., \& Fishman, G. J.  2005, \apj,
625, 60

\bibitem[Penrose(1969)]{Pen69} Penrose, R. 1969, Nuovo Cimento, 1, 252

\bibitem[Piran(2005)]{Pir05} Piran, T. 2005, Reviews of Modern Physics, 76,
     1143

\bibitem[Urry \& Padovani(1995)]{Urr95} Urry, C. M. \& Padovani,
      P. 1995, \pasp, 107, 803

\bibitem[Wald(1974)]{Wal74} Wald, R. M. 1974, \prd, 10, 1680

\bibitem[Zhang \& M\'{e}sz\'{a}ros(2004)]{Zha05} Zhang, B. \& M\'{e}sz\'{a}ros, P.
    2004, Int. J. Mod. Phys., A19, 2385

\end{thebibliography}
\end{document}